\documentclass [12pt,twoside]{article}           
\usepackage[left,modulo]{lineno}
\usepackage{setspace}

\countdef\arxiv=201

\ifnum\arxiv=0 \input cpreb.tex \fi

\ifnum\arxiv=1

\usepackage{epsfig,times,lscape}
\usepackage[usenames]{color}

    \ifnum\count102=1

    \fi

    \ifnum\count102=2
\fi

\newcommand{\nc}{\newcommand}

     \ifnum\count101=1
\nc{\qI}[1]{\section{{#1}}}
\nc{\qA}[1]{\subsection{{#1}}}
\nc{\qun}[1]{\subsubsection{{#1}}}
\nc{\qa}[1]{\paragraph{{#1}}}

\def\qpar{\vskip 2mm plus 0.2mm minus 0.2mm}
\def\qL{\hfill \break}
     \fi 

      \ifnum\count101=2
 \nc{\qI}[1]{\parindent=0mm \vskip 8mm 
{\centerline{\LARGE \color{red}#1}}\vskip 3mm}
\nc{\qA}[1]{\vskip 2.5mm \noindent 
{{\bf\large\color{blue}  #1}} \vskip 1mm \parindent=0mm}
 \nc{\qun}[1]{\vskip 1mm \noindent {\sl #1 }\quad }

\def\qL{\hfill \break}
\def\qpar{\vskip 2mm plus 0.2mm minus 0.2mm}

      \fi

\def\qth{\vrule height 12pt depth 0pt width 0pt}
\def\qtb{\vrule height 0pt depth 5pt width 0pt}

\nc{\qfoot}[1]{\footnote{{#1}}}

      \ifnum\count101=1
\def\qbu{\hfill \par \hskip 6mm $ \bullet $ \hskip 2mm}
\def\qee#1{\hfill \par \hskip 6mm (#1) \hskip 2 mm}
      \fi
      \ifnum\count101=2
\def\qbu{\hfill \par \hskip 4mm $ \bullet $ \hskip 2mm}
\def\qee#1{\hfill \par \hskip 4mm (#1) \hskip 2 mm}
      \fi

\def\qparr{ \vskip 1.0mm plus 0.2mm minus 0.2mm \hangindent=10mm
\hangafter=1}

     \ifnum\count101=1 
 \def\qdec#1{\parindent=0mm\par {\leftskip=2cm {#1} \par}}
     \fi
     \ifnum\count101=2
  \def\qdec#1{\parindent=0mm \par {\leftskip=1cm {#1} \par}}
  
  \def\qcitb#1{\noindent \hbox to 102mm{\hfill \small #1} \vskip 1mm}
      \fi

 \def\qpages#1{\count102=0{\loop\advance\count102 by 1
 \null \vfill\eject \ifnum\count102<#1 \repeat}}

\def\qth{\vrule height 12pt depth 0pt width 0pt}
\def\qtb{\vrule height 0pt depth 5pt width 0pt}

\def\qv{\vskip 0.1mm plus 0.05mm minus 0.05mm}
\def\qhu{\hskip 0.6mm}
\def\qhv{\hskip 3mm}

\def\qhw{\hskip 1.5mm}
\def\qleg#1#2#3{\noindent {\bf \small #1\qhw}{\small #2\qhw}{\it \small #3}\qv }
\fi

\begin{document}
\thispagestyle{empty}



\markboth{{\sl \hfill  \hfill \protect\phantom{3}}}
        {{\protect\phantom{3}\sl \hfill  \hfill}}

\color{yellow} 
\hrule height 20mm depth 10mm width 170mm 
\color{black}
\vskip -1.8cm 

 \centerline{\bf \Large Is cancer a disease that can be cured?}
\vskip 2mm
\centerline{\bf \Large An answer based on a new classification
of diseases}
\vskip 15mm
\centerline{\large 
Peter Richmond$ ^1 $ and Bertrand M. Roehner$ ^2 $
}

\vskip 4mm
\large

{\bf Abstract}\quad Is cancer a disease that can be cured 
or a
degenerative disease which comes predominantly with old
age? We give an answer based on a two-dimen\-sio\-nal
representation of diseases that uses two
parameters to characterize age-de\-pen\-dent changes in
death rates. 
These parameters are defined in the following way.\qL
In mortality curves (extending from
the day after birth to beyond 90 years of age)
there is an age, namely $ a_c \sim 10 $ years, which  
plays a crucial role
in the sense that the mortality decreases in
the interval $ I_1=(a<a_c) $ and increases
in the interval $ I_2=(a>a_c) $. The respective trends in
$ I_1 $ and $ I_2 $ are the two
parameters used in our classification of diseases.
Within the framework of reliability analysis, $ I_1 $ and
$ I_2 $ would be referred to as the ``burn-in'' and ``wear-out''
phases. This leads to define three broad groups of diseases
with respect to $ a_c $.
(AS1) Asymmetry with prevalence of $ I_1 $; 
(AS2) Asymmetry with prevalence of $ I_2 $ and 
(S) Symmetry, with $ I_1 $ and $ I_2 $ both playing
roles of comparable importance.
Not surprisingly, among AS1-cases one finds all diseases due to
congenital malformations and chromosomal abnormalities.
In the AS2-class one finds degenerative diseases,
e.g. Alzheimer's disease and other dementias.
Among S-cases one finds most diseases due to
external pathogens or to a wear-out process.
Cancer is one of those mixed cases and in our representation
it turns out
to be closer to (AS2) than to (AS1).\qL
This representation also provides insight into what we call
an {\it overkill effect} in old age. 
This effect tells us that even a
highly effective cancer therapy would have no influence
whatsoever
on the extent of human life unless {\it all} other diseases
are cured simultaneously.
\qL
We conclude that whilst one might develop a cure 
for many of the known
diseases of old age, new diseases will arise 
and the quest for the ultimate cure
will be never ending. 
Death before the age of 120
seems firmly written in our genetic code
and the best we shall be
able to do is to
help people cope with the inevitability of death as and when it
arises. 

\vfill\eject

\qI{Introduction}

In medical statistics the expression ``infant mortality''
refers to the first year after birth and during this first year 
infant death rates (e.g. early, late and post neonatal) are
defined with respect to live births and not with respect to
the living population at the beginning of the respective time intervals.
The 0-1 year definition has its
origin in the recording process of postnatal mortality data.
However, it is not a logical definition in the sense that
the one-year age interval has no real biological basis.
A more objective definition of
``infant mortality'' is suggested by
reliability analysis. For technical devices, infant mortality
refers to the entire age interval
during which the mortality rate decreases. For humans this
corresponds to the period 0-10 years. It is this definition
of infant mortality which will be used throughout this paper.
\qpar

For technical devices the successive phases correspond to the well-known
stages of the so-called ``bathtub'' curve.
\qee{1} Decreasing failure rate: ``burn-in'' stage during which
defective items fail and are eliminated.
\qee{2} Constant failure rate:
normal service stage characterized by a small number
of random failures.
\qee{3} Increasing failure rate: ``wear-out'' stage
during which failures become more frequent.
\qpar

Fig. 1 suggests that human life comprises similar stages  
with however the qualification that the constant failure phase
covers only a short interval from about 8 to 12 years.
For that reason it will be omitted in the subsequent discussion.
%
\begin{figure}[htb]
\centerline{\psfig{width=17cm,figure=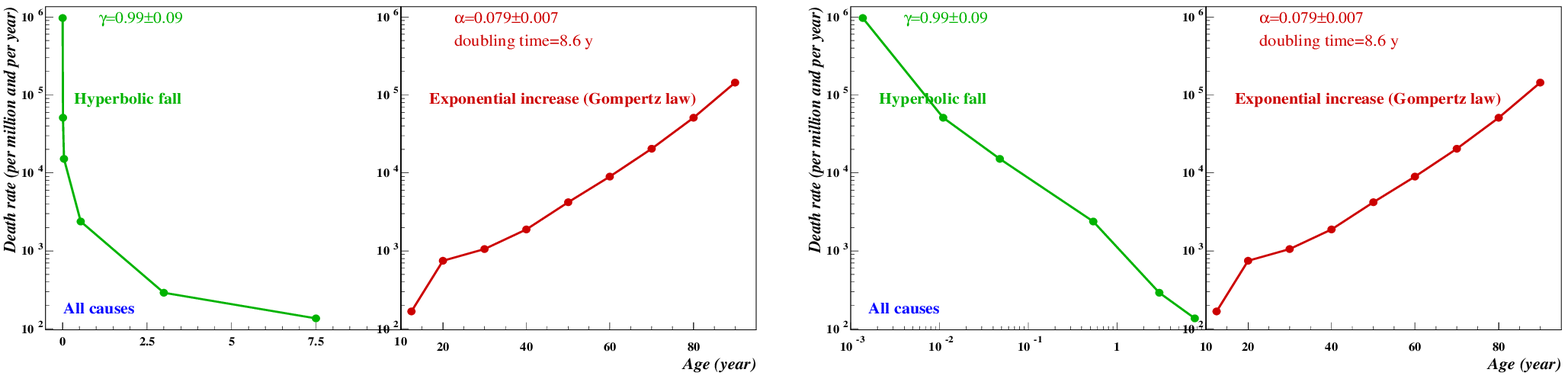}}
\qleg{Fig.\qhu 1a,b\qhv Death rate from all causes,
USA, 1999-2014.}
{In graph (a) the two $ x $ scales are linear 
whereas the $ y $-scales are
logarithmic.
In graph (b) the $ x $ scale  of (0,10) is logarithmic,
whereas the $ x $ scale of (10,90) is linear.
The two straight lines mean that for (0,10) the death rate
is hyperbolic of the form $ y\sim 1/x^{\gamma} $, 
whereas for (10,90) it is an exponential function in accordance
with Gompertz's law $ y\sim e^{\alpha t} $.
It can be noted that due to the huge range of variation
of the death rate, using a linear scale for $ y $ 
would produce a completely useless graph in the sense 
that for $ (0,10) $ it would be a vertical line whereas
for adult deaths it would be
a line superimposed on the $ x $ axis. Thus,
a $ y $ log-scale is not just an option, it is a necessity.}
{Source: CDC ``Wonder'' database 1999-2014
(http://wonder.cdc.gov/ucd-icd10.html).
All data are average annual death rates over the 16-year
long interval 1999-2014.}
\end{figure}

\qpar
The burn-in and wear-out phases are characterized
by two symmetrical processes: elimination of defects in the
burn-in phase and accumulation of defects in the wear-out
phase. 
\qpar

At first sight such a characterization may seem too
crude to be applied to living organisms.
But two observations will help us 
realize that it can capture
essential features of living systems.
\qbu Table 1 shows that deaths linked to perinatal 
problems and congenital malformations
dominate death rates in the first month
after birth. In subsequent months 
the share of these two causes decreases.
Overall, for the first year they represent 70\%
of the deaths, 16\% in the second year
and 6.6\% for the $ 9-10^- $ age group.
%
\begin{table}[htb]

\small

\centerline{\bf Table 1\quad Predominant causes of death in the
infant phase.}

\vskip 5mm
\hrule
\vskip 0.7mm
\hrule
\vskip 2mm

$$ \matrix{
\qtb
 \hbox{Days after birth}\hfill
& 0-1^- &  1^+-7^- &  7^+-28^- & 28^+-365^- \cr
\noalign{\hrule}
\qth
\hbox{Complications of pregnancy and delivery (P00-96)} \hfill & 
79\% & 65\% & 55\% & 7.0\% \cr
\hbox{Congenital malformations and abnormalities (Q00-99)} \hfill & 
18\% & 26\% & 25\% & 17\% \cr
\qtb
\hbox{\color{blue} Total (P00-96 + Q00-99) in percent of ``All
  causes''} \hfill & 
\color{blue} 97\% & \color{blue} 91\% & \color{blue} 80\% &
\color{blue} 24\% \cr
\noalign{\hrule}
} $$
\vskip 1.5mm
\small
Notes: Deaths related to perinatal problems and congenital
malformations are completely predominant in the first month
after birth. Here $ 1^- $ denotes the end of the first day
(i.e. 23h after birth)
whereas $ 1^+ $ denotes the beginning of the second day
(i.e. 25h after birth).
The code numbers following the causes of death are from
the ICD-10 classification, that is to say 
the 10th revision of the ``International 
Classification of Diseases''; in most countries it was introduced
between 1995 and 2000.
\qL
{\it Source: CDC (Centers for Disease Control): ``Wonder''
database, 1999-2014.}
\vskip 5mm
\hrule
\vskip 0.7mm
\hrule
\end{table}
%
\qbu
Many in both the medical community and population at large
seem to overlook the fact that in old age
hiding behind any disease that is more or less curable,
there is another one which 
probably will be more difficult to treat for the simple reason
that it occurs further along the `wear-out'
process. For instance, cancer may be followed by
Alzheimer's disease or another degenerative disease.
Fig. 2 from Courchesne et al. (2000) 
shows that the volume of gray matter
(i.e. the external layer of the brain) decreases steadily
after the age of 10. It is true that the volume of white matter
decreases more slowly but because
good functionality requires a combination of both white and gray 
matter%
\qfoot{One needs only remember that white matter consists of
more than 100,000 kilometers of nerve fibers which connect parts of
the gray matter with each other.}%
, 
a loss of capabilities seems inevitable.
Moreover, the inset shows that instead of being
an ``orderly'' shrinkage,
it is a wear and tear process which
distorts the shape of the brain. Short of freezing the whole
dynamic of aging, it is difficult to imagine how
it can be stopped or reversed. Within such
a picture, the diseases of aging should be seen 
as recurrent manifestations of the global aging mechanism. 

%
\begin{figure}[htb]
\centerline{\psfig{width=10cm,figure=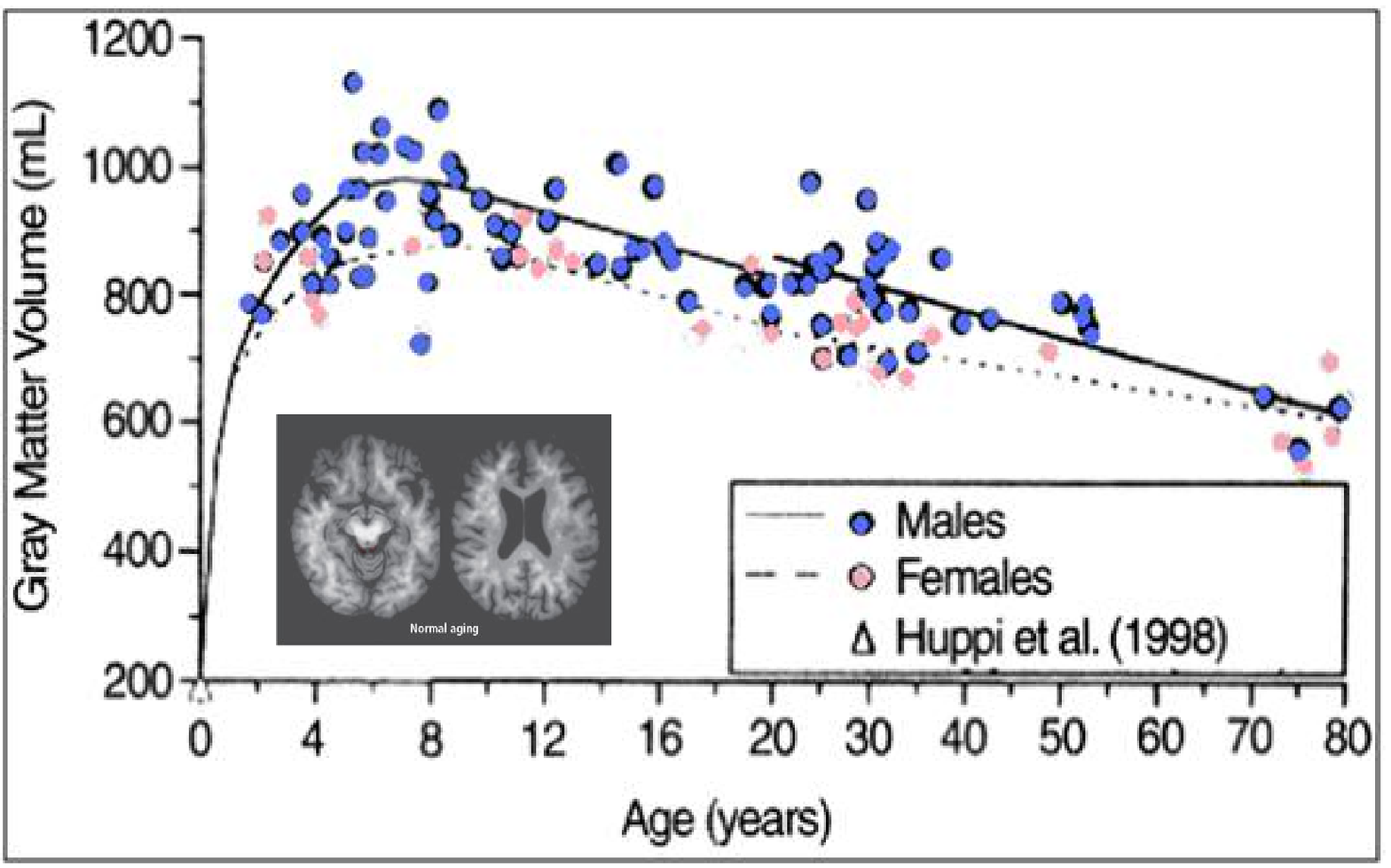}}
\qleg{Fig.\qhu 2\qhv Decrease of the volume of gray matter.}
{The graph describes quantitatively the shrinkage that the
inset shows qualitatively. The data are for 116 healthy
volunteers whose mean age was 21.4 years. From the data points
it is apparent that the sample comprised only few elderly 
people; thus, there was nobody between 50 and 65 or over 80.} 
{Sources: Adapted from Courchesne et al. (2000) 
and http://www.health.harvard.edu/brain-imaging (inset).}
\end{figure}

\qI{Classification of diseases}

Before we define the classification, it should be noted 
that in principle one needs
not only death data but also incidence data.
By relying on deaths we actually mix up two phenomena which should
be studied separately, that is (i) The emergence of a disease
and (ii) How well a disease can be treated. Diseases
which are 100\% curable will not show up
in our data
and those which are almost completely curable will
appear in reduced form. Fortunately two facts contribute
to make death data relevant nevertheless.
\qbu For most diseases curability is rather limited
for the youngest ($ <1 $ day) and oldest ($ >90 $ year)
age groups. This point is related to the so-called
rectangularization%
\qfoot{In our representation it would rather be
a triangularization, see Fig. 1 in which there is no
plateau.} 
of the death rate curve
(see Mesl\'e and Vallin 2002 and also
Berrut et al. 2016, p. 403, 412). 
\qbu We are not interested in absolute death rate values
but only in relative death levels and the latter
are more or less proportional to incidence levels.
When a disease can be completely cured it tends to
disappear (e.g. smallpox, cholera or tuberculosis).

\qA{Class AS1: diseases in which burn-in predominates}

Fig. 3 shows a typical case in which burn-in 
plays a predominant role.

%
\begin{figure}[htb]
\centerline{\psfig{width=12cm,figure=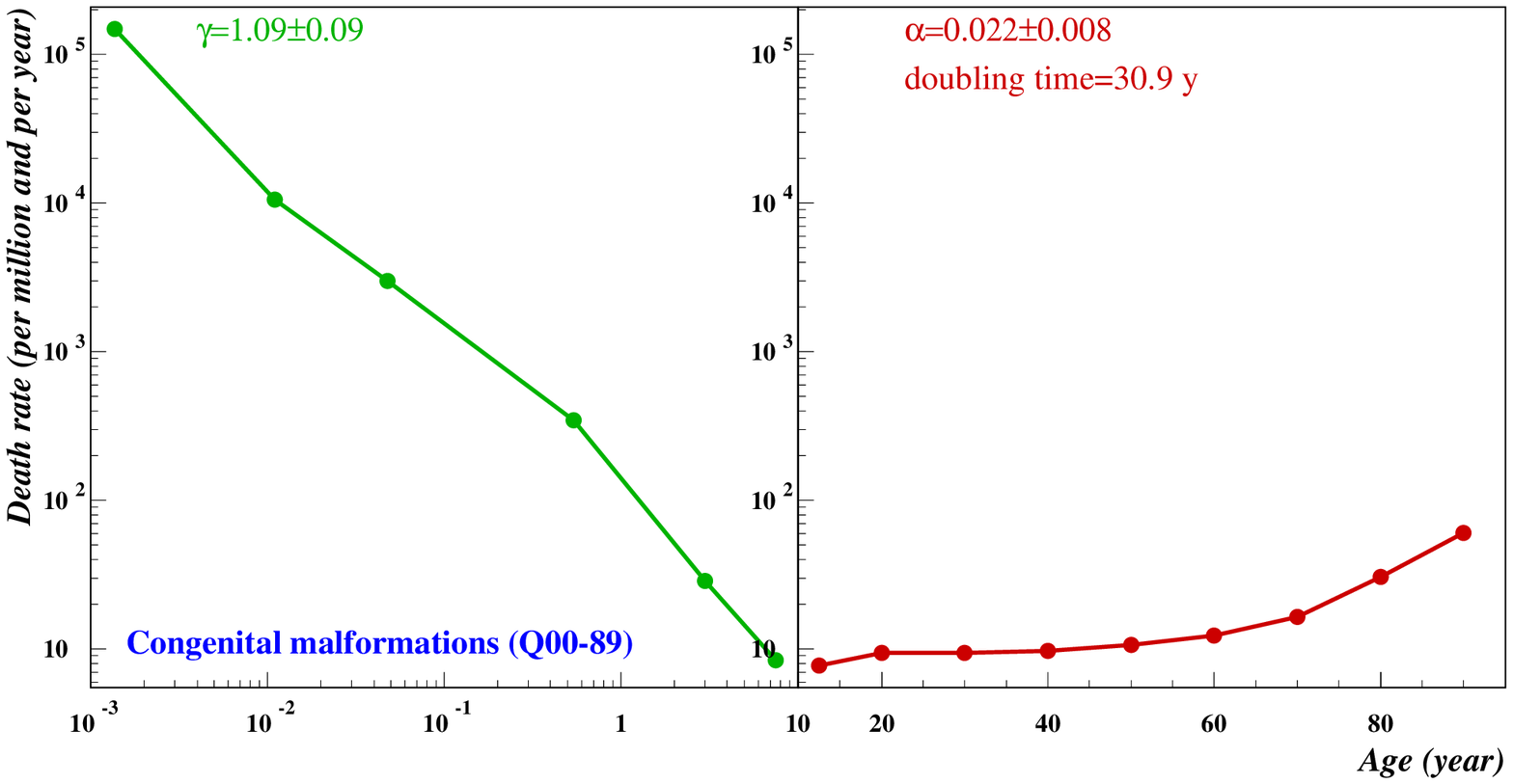}}
\qleg{Fig.\qhu 3\qhv Death rate due to burn-in.}
{On the left, between birth and 10 years,
the death rate is divided by $ 10^4 $ whereas
on the right it is multiplied by a much smaller factor,
namely only 7. This shows that there is a very active 
burn-in process which contrasts with a rather sluggish
wear-out process.}
{Source: Same as in Fig. 1}
\end{figure}

Some congenital malformations may lead to death in the hours
after birth whereas others may bring about death
a long time after birth. 
As a matter of fact, it is quite remarkable that even 50 years
after birth the cause of death can be traced back to a congenital
malformation.
\qpar
Similar values of the exponent $ \gamma $ 
would be obtained for cases of the same kind.
Thus infant mortality due to ``Chromosomal abnormalities'' (Q90-99)
has $ \gamma =1.18 $ and ``Complications of pregnancy and
delivery'' (P00-96) has $ \gamma=1.59 $.

\qA{Class AS2: diseases in which wear-out predominates}

%
\begin{figure}[htb]
\centerline{\psfig{width=12cm,figure=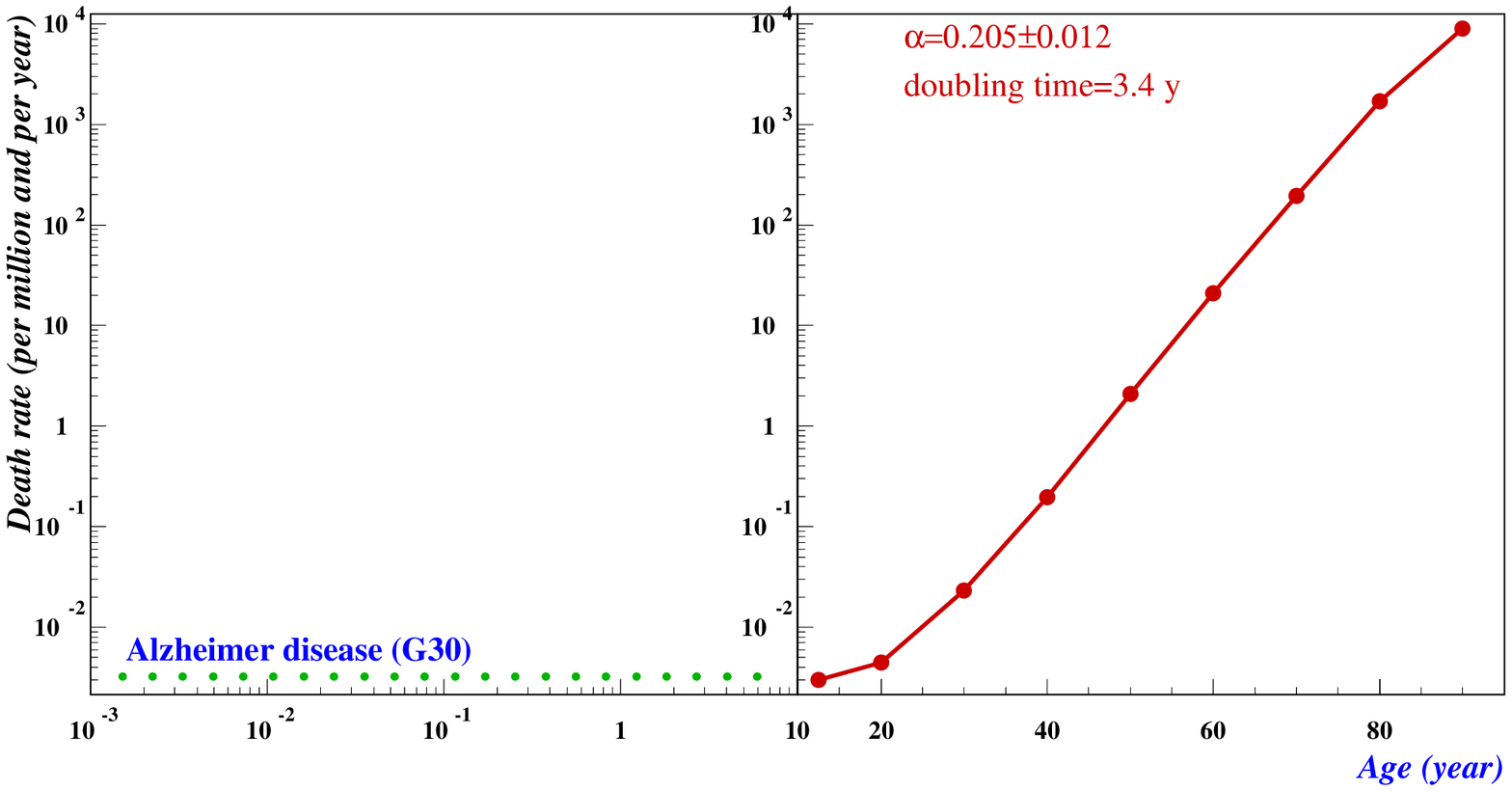}}
\qleg{Fig.\qhu 4\qhv Death rate due to wear-out.}
{The first deaths due to Alzheimer's disease occurred in
the $ 10^+-15^- $ age group; thus the green broken line 
should be seen as
symbolically representing the level zero for younger
age groups.}
{Source: Same as in Fig. 1}
\end{figure}

For the case of Alzheimer's disease shown in Fig. 4 
there is no $ I_1 $ phase
in the sense that for this disease the first death occurs
in the $ 10^+-15^- $ age group. A similar shape is observed
for similar diseases. Thus for all dementia the contribution
of $ I_1 $ is also very small and in the $ I_2 $ phase
the doubling time is 6.8 years.

\qA{Intermediate cases}

By ``intermediate cases'' we mean diseases whose rate decreases
substantially in the infant phase and raises significantly
in the aging phase.

{\it \color{blue} Common features of diseases due to pathogens}\qL
`Bacterial diseases' such as the one
shown in Fig. 5 are typically of this kind.

%
\begin{figure}[htb]
\centerline{\psfig{width=12cm,figure=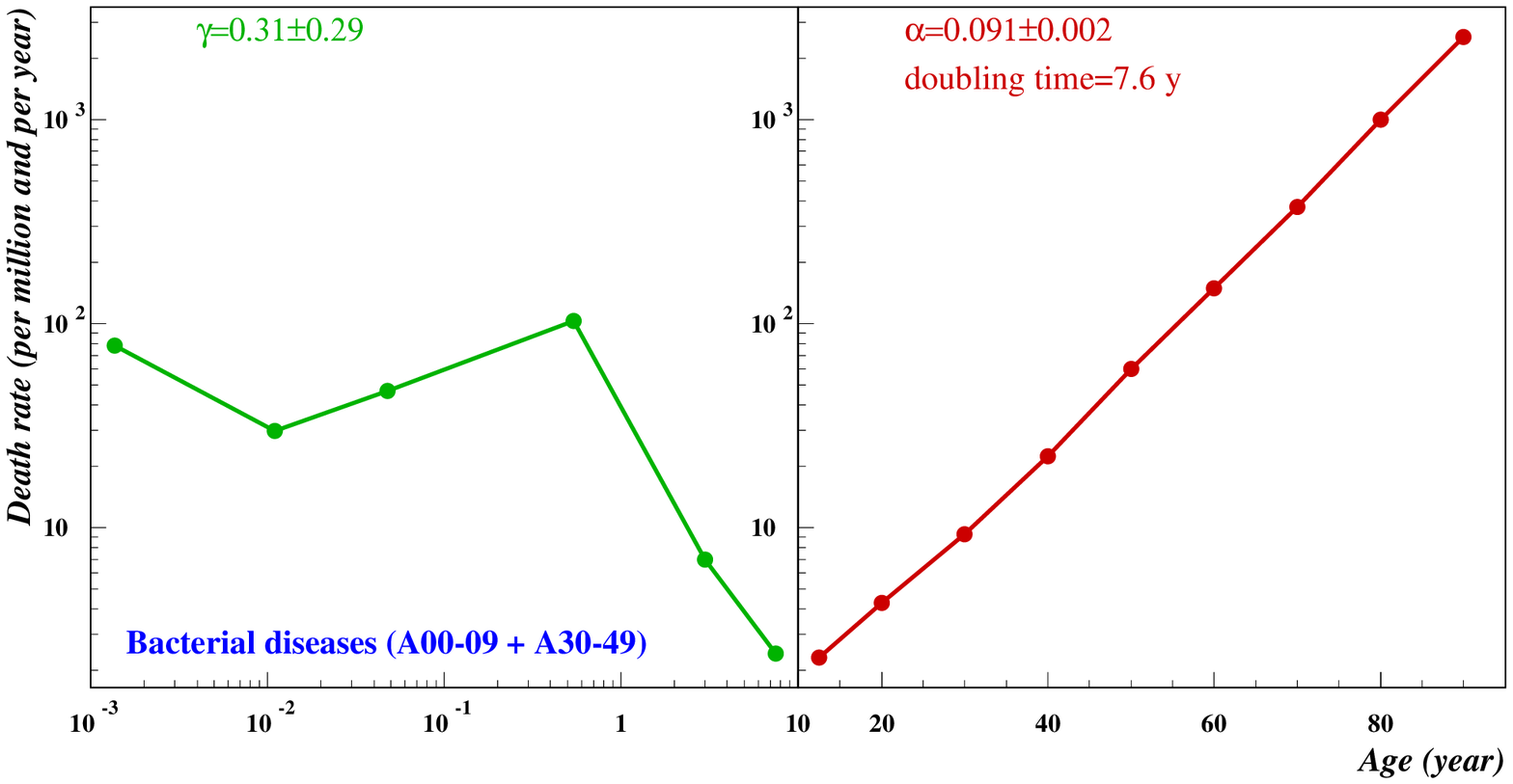}}
\qleg{Fig.\qhu 5\qhv A case
in which there is both burn-in and wear-out.}
{Based on the ranges of the two graphs one may
say that the wear-out process is more considerable than the
burn-in process.}
{Source: Same as in Fig. 1}
\end{figure}

This time, in contrast with Fig. 4, the decline
in the infant phase cannot be interpreted as being due to the
existence of pre-natal defects. 
How then should one interpret such a graph?
A natural interpretation is to say that the decline is due
to a strengthening of the immune system. If correct, this
interpretation would lead us to say that there is a weakening
of the immune system starting right away after the age of 10.\qL
Incidentally, we point out here that the peak which occurs
between the age of one month and one year is not a statistical
fluctuation for it is found in many countries and 
various time periods.
We might speculate
that it arises in a similar manner to that observed for fish as they
leave the yolk sac and finally forage independently. Thus here perhaps
the peak is related to the fact that
between 6 and 12 months, babies shift from the anti-bodies
given by the mother to their own immune system. 
\qpar
This interpretation of the dynamics of the immune system 
can be tested by examining other cases in
which one would also expect the immune system to play a role.
For instance, in table 2 we examine the case of viral infections, 

\begin{table}[htb]

\small

\centerline{\bf Table 2\quad Similarities and differences
between diseases due to bacteria or viruses.}

\vskip 5mm
\hrule
\vskip 0.7mm
\hrule
\vskip 2mm

$$ \matrix{
\hbox{Disease}\hfill & \hbox{Position of} & \hbox{Exponent of}&
\hbox{Doubling time of} \cr
\hbox{}\hfill & \hbox{secondary peak} & \hbox{infant decline}&
\hbox{adult phase} \cr
\qtb
\hbox{}\hfill & \hbox{[year]} & \hbox{}&
\hbox{[year]} \cr
\noalign{\hrule}
\qth
\hbox{Bacterial diseases (A00-09+A30-49)}\hfill & 0.53 & 0.31\pm 0.29& 
7.6 \pm 0.16\cr
\hbox{Influenza and pneumonia (J09-18)}\hfill & 0.53 & 0.31\pm 0.24& 
7.3 \pm 0.70\cr
\qtb
\hbox{Viral infections (A80-B34)}\hfill & 0.05 & 0.52\pm 0.21& 
15.4 \pm 10.5\cr
\noalign{\hrule}
} $$
\vskip 1.5mm
\small
Notes: It should be noted that the 
class J09-18 is dominated by bacterial pneumonia (separate
data for viral versus bacterial pneumonia are shown in Fig. 6a).
Although it follows broadly the same pattern,
the viral case differs
from the bacterial case in two respects (i) the 
secondary peak occurs earlier (ii) the mortality rate
levels off after the age of 50 instead of increasing
steadily as in Fig. 5.
These features remain unexplained so far.
\qL
{\it Source: CDC (Centers for Diseases Control): ``Wonder'' 
database, 1999-2014.}
\vskip 5mm
\hrule
\vskip 0.7mm
\hrule
\end{table}

{\it \color{blue} Difference in adult profiles between bacterial and
  viral diseases}\qL
For viral diseases the increase in the
adult phase is much slower than for bacterial diseases.
In addition, as shown in Table 2, the secondary peak 
is weaker and occurs earlier.

\count101=0  \ifnum\count101=1
\begin{figure}[htb]
\centerline{\psfig{width=14cm,figure=bacvir.eps}}
\qleg{Fig.\qhu 6a,b\qhv Bacterial versus viral diseases.}
{The range of variation for bacterial diseases is 
about 100 times larger than for viral diseases.}
{Sources: Same as for Fig. 1.}
\end{figure}
\fi

\qA{Distribution of diseases in the $ (\gamma, \alpha ) $ plane}

The values of the parameters $ \gamma, \alpha $ estimated previously
for various diseases can now be used to position each disease
in the $ \gamma, \alpha $ plane (Fig. 6).
%
\begin{figure}[htb]
\centerline{\psfig{width=10cm,figure=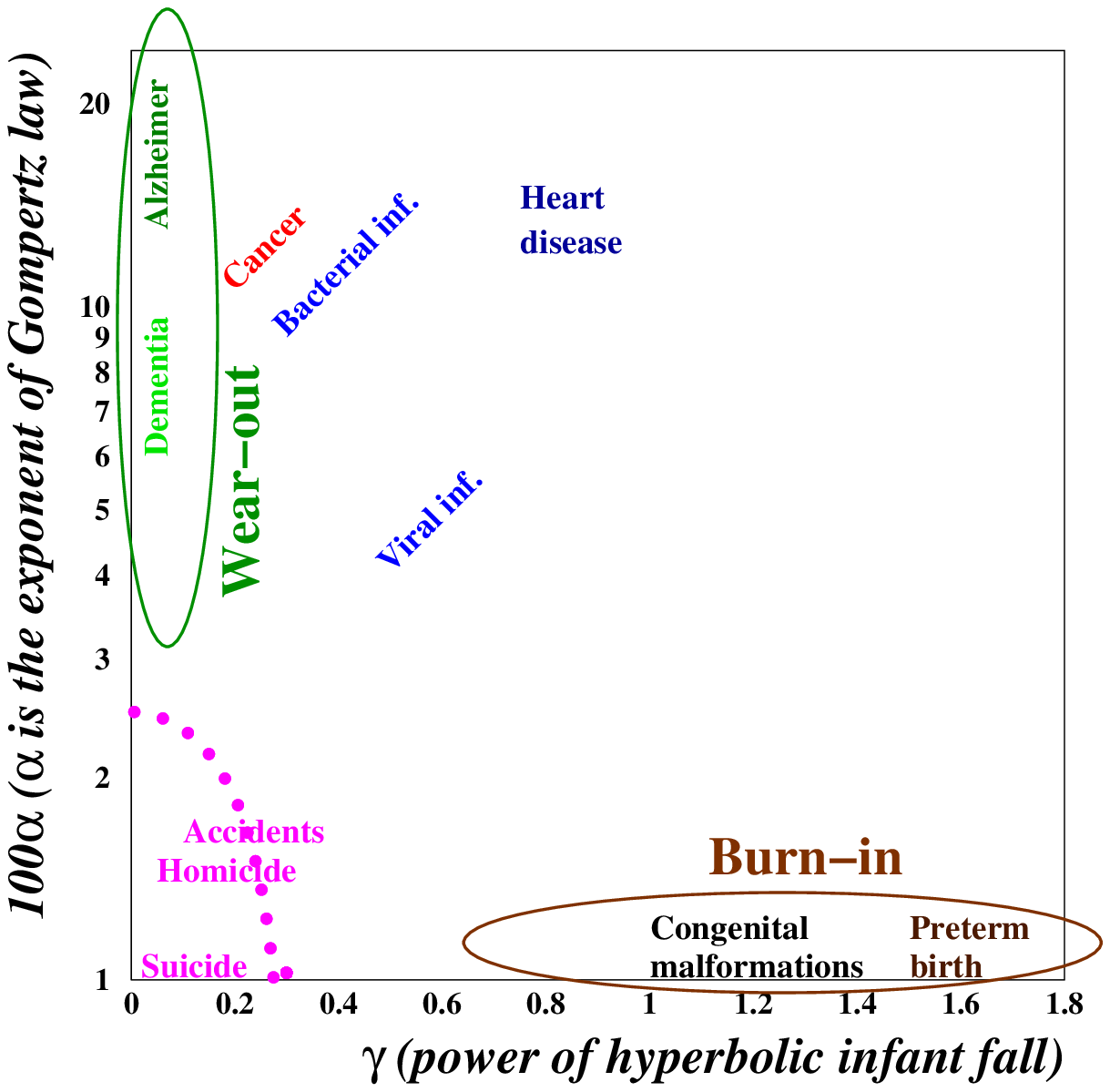}}
\qleg{Fig.\qhu 6\qhv Location of diseases in the $ (\gamma,\alpha) $
plane.}
{$ \gamma $ is the power of the hyperbolic 
fall $ \mu_b \sim 1/t^{\gamma} $
of the infant phase while $ \alpha $
is the exponent of the exponential
increase  in the adult phase: $ \mu \sim e^{\alpha t} $.
Along the $ \gamma $ axis are all diseases related to
congenital defects
including those resulting from preterm birth. 
Along the $ \alpha $ axis
are all degenerative diseases. In between the remaining diseases
may be closer to the first or second group denoting a stronger
congenital or degenerative component respectively.}
{}
\end{figure}

\qA{What is the status of cancer?}

we now discuss the status of cancer in the present
framework.
\qpar

There is a long standing debate about the status of cancer.
Is it a disease that is susceptible of being cured
(like tuberculosis, say) or is it a degenerative disease 
(like Alzheimer's disease, say)? Can our
framework give us some clues?
\qpar

%
\begin{figure}[htb]
\centerline{\psfig{width=12cm,figure=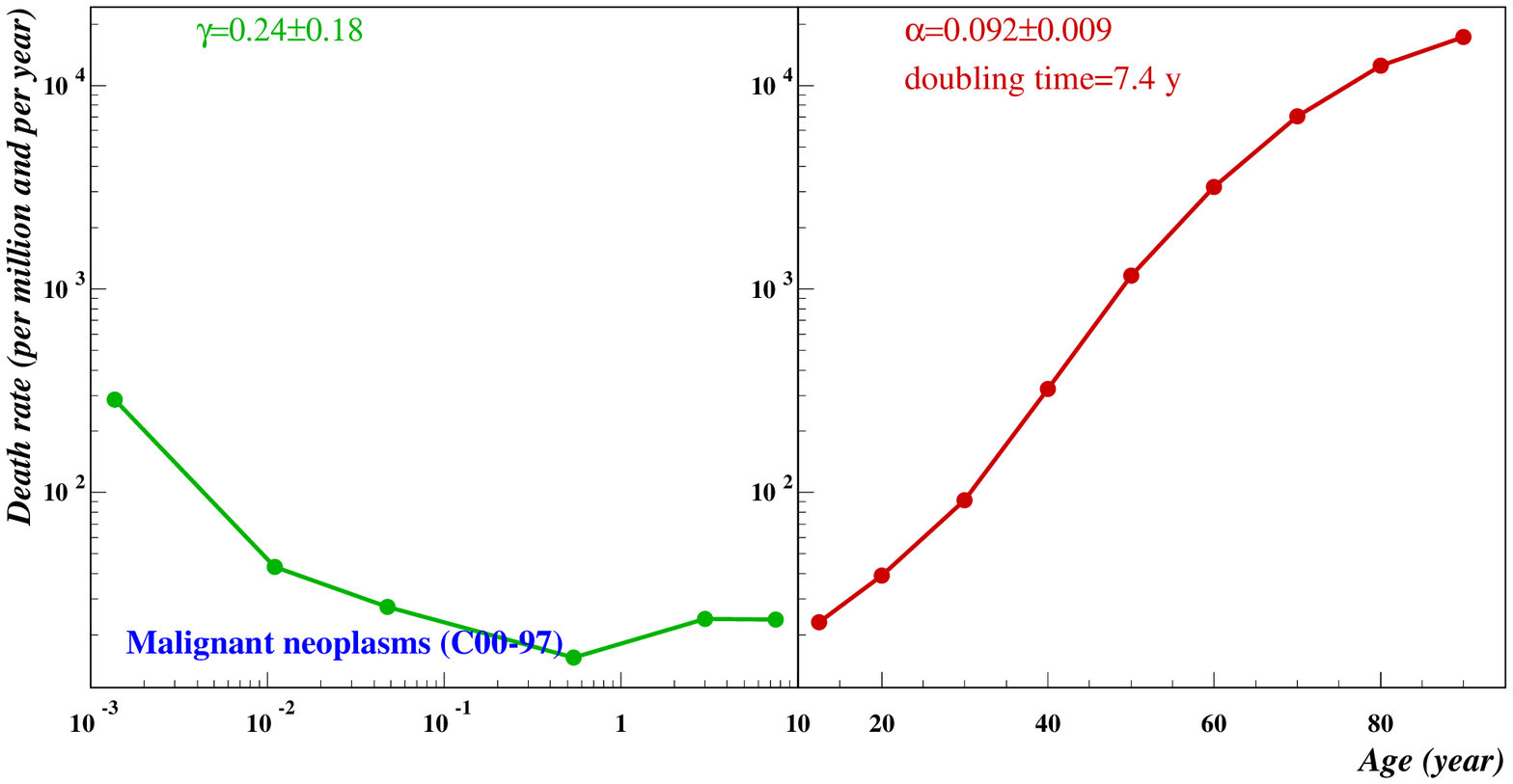}}
\qleg{Fig.\qhu 7\qhv Cancer.}
{Based on the respective ranges of variation 
of the infant and adult death rates
one may
say that the wear-out process is more considerable than the
burn-in process.}
{Source: Same as in Fig. 1}
\end{figure}

For heart diseases the wear-out explanation seems fairly 
natural.
Nobody would claim that it is possible
to cure completely {\it all} heart diseases. 
If one accepts the idea that the heart
is a pump the fact that it is limited to 
a certain number of beats seems reasonable%
\qfoot{Cross species comparisons suggest that this number
is somewhere between one and two billion beats.}%
.
\qpar

{\it \color{blue} Public attitude toward cancer.}\quad 
For cancer the stated objective of many programs is
to find ``a cure for cancer'', to ``beat cancer''
or to ``win the war on cancer''%
\qfoot{Just as an illustration, here are three typical sentences.
\qbu ``Those who took part in the overnight relay
will not rest until a cure for cancer has been found''.
\qbu ``Twelve hours of fun, friendship and fundraising to beat
cancer''.
\qbu 
In his speech after signing the `National Cancer Act' of 1971,
President Richard Nixon used the expression ``the conquest of 
cancer'', perhaps in a parallel with the conquest of the Moon.}%
.
Thus, the goal is not just to find a cure for specific forms of
cancer but to eradicate cancer as a major cause of death,
just as tuberculosis was eliminated as a major disease.
\qpar

{\it \color{blue} Status of cancer in the present classification.}\quad
If cancer is due to a few mechanisms that may be
knocked out, a global remedy may indeed be conceivable.
However, if it is due to a (programmed) weakening of the 
immune system that hope does not seem realistic.
One faces the same problem for technical devices in the sense 
that it is not always obvious whether an increasing failure rate 
is due to one isolated hidden defect or to the simultaneous attrition
and wear of many components. 
In the first case the defect may be
eliminated, however in the second case one would need to replace almost
every part of the device. \qL
Fig. 7 shows that in
the infant phase cancer looks very similar to 
the case of bacterial diseases of Fig. 5 except that there
is no secondary peak. In their adult phases the two cases
look also very similar both in shape and doubling-time.
The case of bacterial diseases
was (tentatively) explained by successive changes in the strength
of the immune system.
Surprisingly for a disease such as tuberculosis for which
a treatment is available, its adult phase mortality
has a doubling time of about 9.1 years which is not
very different from the doubling time of cancer.
The same observation holds for other bacterial diseases for which
antibiotics are supposed to be highly effective.
These features are consistent with a conception
in which one assumes that any therapy, no matter how effective,
can only 
help the immune system but cannot replace it. Thus, even 
for a well-treated disease such as tuberculosis, the therapy
barely influences the global death rate at ages beyond 100.

\qI{Can the human life span be extended?}

Over the past century the life expectancy at birth
in developed countries has been multiplied at least
by a factor of two but this was achieved through a fall
in infant and middle-age adult mortality.
The ultimate boundary of human life span is still
around 120 years (for more detail 
in this respect see Richmond et al. 2016).
Thus, a natural question is whether the life 
span can be extended and what are the conditions
for that.

\qA{Keeping the meaning of Gompertz's law in mind}

The difficulty of the challenge can be assessed
through the following well-known
(but often forgotten) observation. 
The number $ s(t) $ of  radioactive atoms in
a sample 
decreases as an exponential function of their ``age'':
$ s(t)=e^{-\mu t} $. ``Beating'' an exponential function
is already a difficult task; yet, this decrease corresponds
to a constant death rate $ (1/s)ds/dt=\mu $ whereas Gompertz's law
tells us that in a population of humans the death rate itself
increases exponentially with age $ \mu(t) \sim e^{\alpha t} $.
The graphs presented in the first part of this paper show
that the same law also holds at the level of most individual
diseases, albeit with different values of $ \alpha $.
Thus, one needs to ``beat a double exponential''
which is quite a challenge.
\qpar

It is true that this fairly standard argument 
may not sound very compelling
because it is purely mathematical; that is
probably why it is often overlooked.
In what follows we present
it in a more concrete way which, at the same time,
will reveal the conditions that must be fulfilled 
to ``beat'' Gompertz's law. 
\qpar
First, we introduce a classification
of diseases based on how effectively we can treat them.

\qA{Differentiated responses of diseases to therapy}

Diseases respond in different ways to our efforts
to cure them. This suggests a distinction
between the following classes.
\qee{1}  Diseases which have been eradicated in 
developed countries,
e.g. smallpox, rabies, cholera%
\qfoot{It can be remembered that as late as 1892
an outbreak of cholera claimed 8,600 deaths in Hamburg.
Throughout the 19th century there had been similar
cholera epidemics in London, New York and Paris.}%
, 
leprosy. In terms of age-specific death rates
these diseases have a rate close to zero
(say below 1 per million) even in old age.
\qee{2} Diseases which are more or less curable, e.g.
tuberculosis, pneumonia, hepatitis.
These diseases are characterized by death rates
which are of the order of 1 per million at age 30
but show a Gompertz-like increase pattern which leads to
substantial rates (of the order of 1,000 per million)
in old age.
\qee{3} Diseases which are not yet curable,
e.g. cerebrovascular accidents,
multiple sclerosis, Alzheimer's disease.
Like those of (2) these diseases have a low rate 
in young adults but not at all for the same reason.
In (2) it is because an effective therapy is available
whereas in (3) it is because the wear-out process 
has only just started.
\qpar
The diseases in the first two classes are caused
by exogenous pathogens whereas those in class (3)
are mostly due to endogenous causes.
Cancer and heart diseases may lie somewhere between (2) and (3)
but closer to (3) than to (2).
\qpar
Although different in terms of therapy, the classes (2) and
(3) can be merged together for the purpose of the forthcoming
argument because they both have 
a substantial rate in old age.

\qA{The ``overkill'' argument}
During the Cold War the expression ``overkill capability''
referred to the fact that both the US and the USSR
possessed enough nuclear weapons 
to destroy one another many times over.
Similarly,
mortality in old age is ruled
by an overkill effect. What we mean by that
expression is that even if a complete cure could be found
for 99\% of the diseases with only one disease
remaining unchecked,
this single disease would obliterate 
any sample population, no matter how large, between the
ages of 90 and 120. This is due
to the fact that the decrease in the number
of survivors brought about
by any unchecked disease is exponential which implicates
that, unless
the exponent is very small (which would correspond to
diseases that are eradicated), such a decrease will
suppress any population within a few years.
Just as one atomic bomb may destroy a whole city, similarly
a single unchecked disease may reduce a whole population
to extinction. 
\qL
Let us illustrate this argument by the case of cancer.
\qpar
According to Fig. 8d, presently the death rate from
cancer at the age of 90 is $ \mu(90)=1 $ per 100 and per year.
This means that in a sample of 100 people aged 90
the number of survivors will decrease as:
$$ (1/s)(ds/dt)=\mu \rightarrow s(t)=s(90)\exp[-\mu (t-90)]=
100\exp[-1\times (t-90)] $$
where $ t $ is the age.
Thus, one gets: 
$$ s(100)=100\exp(-1\times 10)=0.0045,\quad s(120)\simeq 10^{-13} $$
Thus, a single unchecked disease is
able to suppress the whole population
within a few years. \qL
It can be observed that this is
a very conservative calculation in the sense that we assumed
that $ \mu(t) $ keeps its $ \mu(90) $ value in
subsequent years whereas in fact for almost all diseases
$ \mu(t) $ continues to rise after 90 even though possibly
at a slower rate than before 90.
\qpar
In contrast, for an eradicated disease the
mortality at 90 is of the order of $ 10^{-4} $ per 100.
Under such conditions
the previous
calculation would read:
$$ s(100)=100\exp(-0.0001\times 10)=99.9,\quad 
s(120)=99.7,\quad s(200)=98 $$

In this case there is a marked extension of the human life span.
However, as stressed above, 
for this impressive effect to work
not a single unchecked disease should remain.
\qpar

The overkill argument can be checked by observation.
Over the past century the eradication of several diseases
together with
better treatment for a number of others led
to a marked increase in the number of centenarians and
supercentenarians (i.e. older than 110).
Yet, the upper bound of the human life span did not
change. In 1997 Jeanne Calment died at the age of 
122 years and 164 days, a record which remains 
unmatched ever since.
Moreover she was, so to say, an outlier
in the sense that the second oldest person reached
an age of only 119 years and 97 days%
\qfoot{The fact that most of the oldest persons are
living in developed countries does not mean that there
are no supercentenarians in developing countries. It is
rather due to the fact that in the early 20th century
in many developing countries birth dates were not recorded 
in a systematic and reliable way. The same reason explains
that there are no reliable data for supercentenarians
in the 18th or 19th centuries. Of course, 
reliable birth data would exist for persons belonging
to the aristocracy but such persons 
represented but a small proportion
of the total population.}%
.
Even if in the future the record of 122 years is broken
most likely it will be surpassed by only a short margin
of less than one year.

\qI{How would an effective cancer therapy work in old age?}

In 1900 pneumonia and tuberculosis were the leading causes of
death in the United States with annual rates of the order 
of 2,000 per million%
\qfoot{At that time cancer came only in 8th position.}%
.
However, by 1950 death rates had decreased by a factor of 10
and further reductions occurred in subsequent years.
Thus, such therapies can be said as having been
really effective. What was their effect in old ages?
In other words
how did this therapy affect the shape of the
age-specific curves. Before performing the test we were
expecting that, in accordance with Gompertz's law,
the exponential shape would remain unchanged.
This is indeed what we observed 
albeit with a notable increase in the exponent.
In short, even such highly effective therapies had
a much reduced effect on old age mortality. The same
outcome is likely for cancer.

%
\begin{figure}[htb]
\centerline{\psfig{width=13cm,figure=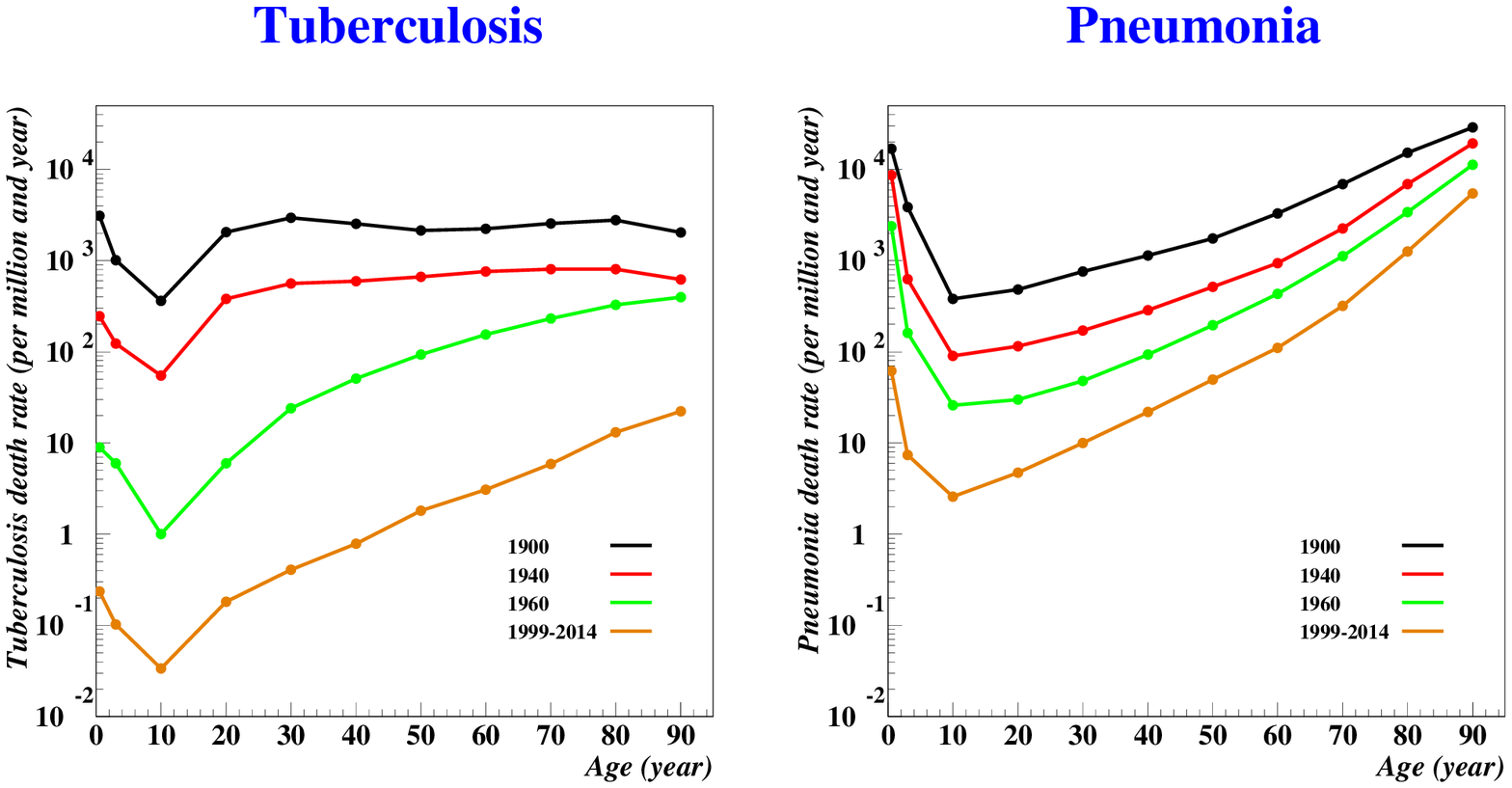}}
\qleg{Fig.\qhu 8a,b\qhv Pattern of change of death rates
over one century in the United States.}
{For tuberculosis and pneumonia there is a well-defined pattern
in the sense that in the adult phase
the slope (that is to say the exponent $ \alpha $)
increases as the global mortality decreases. The fact
that for Switzerland almost identical curves are observed
suggests that this pattern is shared also
by other developed countries.
The ICD10 code definition of
tuberculosis and pneumonia are A16-19
and J09-18 respectively.}
{Sources: Linder and Grove (1947, p. 248-254);
Grove and Hetzel (1968, p. 378-469); ``Wonder'' database of the
``Center for Diseases Control'' (CDC).}
\end{figure}

%
\begin{figure}[htb]
\centerline{\psfig{width=13cm,figure=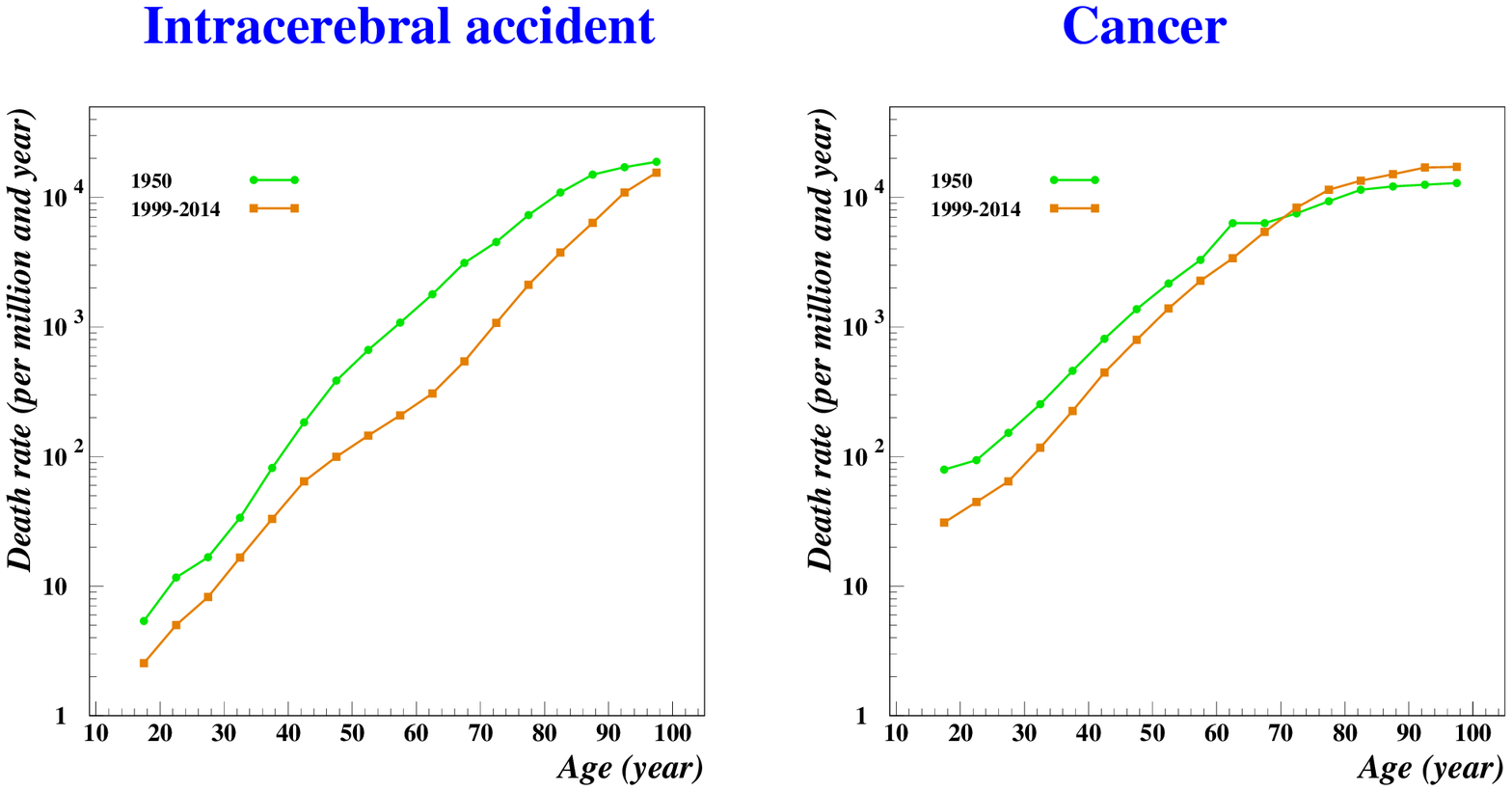}}
\qleg{Fig.\qhu 8c,d\qhv Change of death rates
over 60 years in the United States.}
{With their limited death rate reductions, these
diseases belong somewhere between 
classes 2 and 3 (defined in the text) but
closer to the latter than to the former.
The ICD10 codes for intracerebral accident and cancer
(malignant neoplasms) are I60-64 and C00-97 respectively.}
{Sources: Vital Statistics of the United States 1950, Volume 3;
``Wonder'' database of the ``Center for Diseases Control'' (CDC).}
\end{figure}

The case of pneumonia suggests that the curves 
for different years converge toward
a common limit around 110 years.
This is another instance of a fixed point mechanism
of the kind that we already studied for global mortality
(Richmond et al. 2016). Adopting this viewpoint, the increase
in slope is an obvious consequence since if the
mortality at age 10 falls yet the mortality at age
110 remains stable, the slope of the straight line
joining these two points must increase. For tuberculosis the
convergence to a fixed point (assuming one exists) is
much slower. The fact that the curves for 1900 and 1940 are almost
flat suggests that the immune system barely changed in effectiveness
over time. This might seem at variance with the suggestion that
its effectiveness declines with age. 
But note that the death rate in 1900
is already very high at age around 20 years 
and comparable with that in old
age for adults today. This then suggests that over time lifestyles,
nutrition and medical care have actually improved the effectiveness of
the immune system in young adults today such that the death rate on
the log scale is reduced by 4 orders of magnitude.

\qA{Noteworthy observation about the status of Gompertz's law}

In order to prevent any confusion we wish to emphasize
two points regarding the role played by Gompertz's law in the
arguments developed in the present paper.
\qbu Our arguments require only what may be called
a weak form of the law of Gompertz, basically that the 
death rate increases with age but with a slope
which is allowed to fluctuate with age. For instance,
whether or not the slope decreases in old age%
\qfoot{There has been a debate about this point in the past two
decades.}
is irrelevant for our reasoning as can clearly 
be seen in the subsection about the overkill effect.
In fact, this effect will exist (along with the resulting life span
wall around age 120) even if the death rate is assumed
to be constant beyond the age of 90.  It could even be allowed
to decrease slightly (although this is never observed)
without the conclusion of the argument being changed.
\qbu The global ``all causes''
death rate is
an aggregate variable whose properties depend upon
the underlying components. As the underlying diseases do
not behave in the same way 
their addition will produce a mixed bag which will basically
be controlled by a few predominant components.
We have already observed that the relative importance
of diseases has greatly changed over the past century.
Therefore one should not be surprised to see the 
shape of the global death rate change in the course of
time. Depending on whether tuberculosis or cancer is the 
predominant cause of death the shape will 
not be the same and this
will be all the more visible in old age when the
the age groups become smaller and the fluctuations 
therefore larger.
There may even be transversal
differences from country to country
due to differing relative weights of diseases.
In other words, from a global perspective, 
any discussion about the {\it detailed} shape
of Gompertz's function is irrelevant and fairly useless.

\qA{Conclusion}

This paper could have been entitled ``Is cancer a disease
that can be cured? And, besides, does it really matter?''.
To the first question we could only give an answer in terms
of likelihood. Although we decided against including the second
question into the title, 
it is in fact this second interrogation that
we were able to answer in a definite way.
\qpar

We conclude that whilst one might develop 
a cure for many of the known
diseases of old age, new diseases will arise and 
the quest for the ultimate cure
will be never ending. Death is inevitable and the best we shall be
able to do is to
help people cope with the inevitability of death as and when it
arises. 
\qpar

Naturally, our reasoning 
would be invalidated if
major changes to human DNA are allowed.
But that would open uncharted 
territory which has major ethical implications for society.

\qI{Appendix A: Span of life versus life expectancy}

In the previous section we focused on maximum life span 
rather than on life expectancy for the simple reason
that the 120 year limit is a certainty whereas life expectancy
is a statistical average which gives only a likelihood. 
\qpar

Yet, because life expectancy is a standard demographic
indicator it may be of interest to know how many years
a successful cancer therapy would add to it.\qL
We will give two answers.
%
\begin{table}[htb]

\small

\centerline{\bf Table A1\quad How life expectancy is affected
by a mortality reduction in different age groups}

\vskip 5mm
\hrule
\vskip 0.7mm
\hrule
\vskip 2mm

$$ \matrix{
\hbox{Hypothesis}\hfill & \hbox{Age group 1} & \hbox{Age group 2}&
\hbox{Age group 3} & \hbox{Life expectancy} & H_i-H_0\cr
\qtb
\hbox{}\hfill & 0-5 & 50-55 & 80-85 \cr
\noalign{\hrule}
\qth
H_0: \hbox{ Initial death rates}\hfill & 30 & 30& 940 & 79.20 & \cr
H_1: \hbox{ Reduced death rate in 1}\hfill & 10 & 30& 960
& 80.80 & 1.60 \cr
\qtb
H_2: \hbox{ Reduced death rate in 2}\hfill & 30 & 10& 960&
79.80 & 0.60\cr
\noalign{\hrule}
} $$
\vskip 1.5mm
\small
Notes: All rates are per 1,000. 
In order to better show the principle of the calculation
the number of age groups was limited to 3.
Under hypothesis  $ H_1 $ short lives
of less than 5 years are replaced by lives lasting beyond 80
whereas under hypothesis $ H_2 $ 
medium lives in the interval 50-55 are similarly replaced by
lives of age group 3. Thus, there is no surprise in the fact
that the increase in
life expectancy (i.e. average duration of life) is 
higher under $ H_1 $
than under $ H_2 $. This effect would be even stronger if
group 2 would be an older age group. For instance,
the same calculation with 70-75 as age group 2 gives
a difference $ H_1-H_2 $ equal to 1.40 year instead of 1.00.\qL
This calculation shows that an effective cancer therapy would
affect life expectancy at birth much less than the huge 
infant and early childhood mortality
reduction that occurred during the 20th century.
{\it }
\vskip 5mm
\hrule
\vskip 0.7mm
\hrule
\end{table}
%
\qbu The first answer is taken from papers by Jay Olshansky
and associates (1990, 2001). Their conclusion can
be summarized as follows.
\qdec{Eliminating all forms of cancer would increase life
expectancy at birth by only three years.}
The problem with these papers is that they
rely almost purely on
statistical projections. This gives readers precise 
information but
leaves them without real
understanding. Moreover, these projection
techniques are pushed too far sometimes.
Just as an illustration, 
the paper of 2001 gives projections that extend to 2577;
one wonders how reliable such long-term projections
can be.
\qbu To explain the smallness of the 
3 year extension, the key-point is that
it is much more effective to reduce
infant and childhood mortality than mid-
or old-age mortality. This is a simple statistical effect 
that is illustrated in Table A1. \qL
As in developed countries
infant mortality is currently at a level of 3 per 1,000
(compared to 150 per 1,000 some 150 years ago)
one cannot gain much on this side. Thus, further
reduction must come from therapies for old age
diseases. Not only is this much more difficult to
achieve from a medical perspective because these are
degenerative diseases,
but in addition,
even if successful, it will bring about only modest
extensions of life expectancy.

\vskip 5mm
{\bf References}

\qparr
Beirrut (S.), Pouillard (V.), Richmond (P.), Roehner (B.M.) 2016:
Deciphering infant mortality. 
Physica A 463, 400-426.

\qparr
Courchesne (E.), Chisum (H.J.),
Townsend (J.), Cowles (A.), Covington (J.), 
Egaas (B.), Harwood (M.), Hinds (S.),Press (G.A.) 2000:
Normal brain development and aging: quantitative
analysis at in vivo MR imaging in healthy
volunteers. 
Radiology, volume 216. issue 3.

\qparr
Linder (F.E.), Grove (R.D.) 1947: Vital statistics rates in the
United States 1900-1940. US Government Printing Office,
Washington DC.

\qparr
Grove (R.D.), Hetzel (A.M.) 1968:  Vital statistics rates in the
United States 1940-1960. US Government Printing Office,
Washington DC.

\qparr
Mesl\'e (F.), Vallin (J.) 2002: Mont\'ee de l'esp\'erance de vie
et concentration de l'\^age au d\'ec\`es.
Institut National d'Etudes D\'emographiques, Working paper No 108.

\qparr
Olshansky (S.J.), Carnes (B.A.), Cassel (C.) 1990:
In search of Methuselah: estimating the upper limits to human longevity.
Science 250, 4981, 634-640.

\qparr
Olshansky (S.J.), Carnes (B.A.), D\'esesquelles (A.) 2001:
Prospects for human longevity. 
Science 291, 5508, 1491-1492.

\qparr
Richmond (P.), Roehner (B.M.) 2016: Predictive implications
of Gompertz's law. 
Physica A 447,446-454.

\end{document}